\documentclass[runningheads]{llncs}
\usepackage{graphicx}
\usepackage[numbers]{natbib}
\usepackage{aas_macros}
\usepackage{hyperref}
\usepackage{xcolor}
\begin{document}
%
\title{TCuPGAN: A novel framework developed for optimizing human-machine interactions in citizen science}
%
%
\author{Ramanakumar Sankar\inst{1}\and
Kameswara Mantha\inst{1}\and
Lucy Fortson\inst{1}\and
Helen Spiers \inst{4}
Thomas Pengo \inst{1} \and
Douglas Mashek \inst{1} \and
Myat Mo \inst{1} \and
Mark Sanders \inst{2} \and
Trace Christensen \inst{2} \and
Jeffrey Salisbury \inst{2} \and
Laura Trouille \inst{3}
}
%
%
\institute{University of Minnesota, Twin Cities, Minnesota, USA\and
Mayo Clinic, Rochester, Minnesota, USA \and
Adler Planetarium, Chicago, Illinois, USA \and
The Francis Crick Institute, London, UK}
%
\maketitle              
\begin{abstract}
    
    In the era of big data in scientific research, there is a necessity to leverage techniques which reduce human effort in labeling and categorizing large datasets by involving sophisticated machine tools. To combat this problem, we present a novel, general purpose model for 3D segmentation that leverages patch-wise adversariality and Long Short-Term Memory to encode sequential information. 
    Using this model alongside citizen science projects which use 3D datasets (image cubes) on the Zooniverse platforms, we propose an iterative human-machine optimization framework where only a fraction of the 2D slices from these cubes are seen by the volunteers. We leverage the patch-wise discriminator in our model to provide an estimate of which slices within these image cubes have poorly generalized feature representations, and correspondingly poor machine performance. These images with corresponding machine proposals would be presented to volunteers on Zooniverse for correction,    
    leading to a drastic reduction in the volunteer effort on citizen science projects. We trained our model on $\sim 2300$ liver tissue 3D electron micrographs. Lipid droplets were 
    segmented within these images through human annotation via the `Etch A Cell - Fat Checker' citizen science project, hosted on the Zooniverse platform. 
    In this work, we demonstrate this framework and the selection methodology which resulted in a measured reduction in volunteer effort  by more than $60\%$. We envision this type of joint human-machine partnership 
    will be of great use on future Zooniverse projects. 
\end{abstract}
\section{Introduction}
Improvements in data collection methodology such as robotic surveys in astrophysics, or electron microscope advancements in bio-medicine, have led to an unprecedented growth in data production rates \cite{Wright2019,Spiers2021}. The analysis of much of these data still requires human effort. Hence, to efficiently analyze these large datasets, it is critical to consider leveraging human-machine optimization frameworks, such as active learning strategies with human-in-the-loop ideologies \cite{Fortson2018, Walmsley2022}. One possible framework is to (1) use a machine model to provide an initial analysis of the data; (2) identify poorly performing data; (3) present this subset of data to humans for correction; and (4) use the human-corrected data to refine the machine. Such a framework would enable fast processing of large amounts of data, speeding the analytical pipeline and ensuring more efficient application of human effort. However, one of the greatest challenges in implementing such a framework is the accurate identification of data with poor machine performance, particularly when ground truth labels are unavailable. While uncertainty prediction has become a strong component of new machine/deep-learning studies \citep{Grote2023}, most existing techniques work well for classification tasks \citep{Chua2023}. Generative models are much more difficult due to their more complex architectures and training regiment and would likely require more sophisticated training \citep{Arikan2023}. In this work, we focus on using ad-hoc methods within the generative model (primarily the learnt adversariality), as a proxy for machine uncertainty. 



In this study, we use data from the Etch A Cell - Fat Checker project\footnote{\url{https://www.zooniverse.org/projects/dwright04/etch-a-cell-fat-checker}}, which is a cell biology citizen science project hosted on Zooniverse.org to identify lipid droplets in electron micrographs. Using this project as a case study, we show how machine models can be used to predict which data may benefit from human analysis. 
To satisfy the dual role of improving the completion rate of the project by using machine models to classify large fractions of the data and to determine poorly performing data to present to volunteers, we present the Temporal Cubic Patch Generative Adversarial Network (TCuPGAN), a generative model that works on image cubes (3D images, like volume electron microscopy data) or video data to do 3D segmentation. We apply our model to data from the Etch a Cell - Fat Checker project and present 
our results from testing our framework on this dataset.
In this study, we use the discriminator component as a method for identifying poorly performing data (i.e., those that generalize poorly between the input image cube and the target segments). Our selection method can identify either image cubes or individual slices which show poor machine performance; these can be shown to the volunteers for refinement, thereby significantly reducing both the project completion time and the volunteer effort.
\section{Methods}

We demonstrate our work on the TCuP-GAN, which is based on a generative Image-to-Image translational model \cite[PatchGAN; ][]{isola2017image}. 
PatchGAN is 2D image segmentation model, which uses the U-Net architecture, 
and features both a generative and discriminative component. 
In our model generator, we replace the 2D convolutional layers in the PatchGAN architecture with 2D convolutional Long Short-Term Memory (LSTM) layers to simultaneously capture both the 2D spatial information in each slice as well as the correlations along the third dimension (depth axis in image cubes, or time axis for videos). The LSTM layers extract both the feature vectors ($h_t$, which encodes the image features as a function of depth/time) and the cell state ($c$, which captures the cumulative spatial correlation of the features across the depth/time axis) \cite{Shi2015}. 
Subsequently, the learnt features encode both temporal (or depth-wise) and spatial correlation information. We skip both the $h$ and the $c$ vectors across the bottleneck from the encoder to the decoder as part of the U-Net architecture. 


The discriminator is a patch-wise binary classifier that takes a concatenation of the input image and its corresponding ground truth or generated mask and outputs a $8\times8$ probability matrix per depth slice. The discriminator is a series of 3D convolutions with kernel size $(1, 3, 3)$, followed by Layer Normalization. We use the LeakyReLU activation with $\alpha = 0.2$ for all layers except the final convolutional layer, which has a Sigmoid output on a single channel. Each unit of the output of the discriminator represents a patch of the input image, and provides the probability that the patch is real. 
For each slice, we take the mean and variance of its discriminator scores for our test metrics.


\subsection{Etch-a-Cell: FatChecker data}\label{sec:data}
To train the model, we use data from Etch a Cell - Fat Checker where volunteers are presented with individual 2D slices and asked to annotate the lipid droplets. We aggregate the volunteer responses \citep{Spiers2021} and build the image cube by stacking the 2D images and their corresponding masks.
This Zooniverse project presents $1200\times 1200$ pixel sized 2D image slices to the volunteers, who annotate the outline of the lipid droplets. For each 2D slice, a binary consensus mask is generated by aggregating the annotations of multiple volunteers \citep{Spiers2021}, where pixels corresponding to lipid regions are indicated by $1$ and background is $0$. In this work, we stack these different 2D slices to generate their corresponding 3D image cubes with sizes $1200\times 1200 \times 10$. We then augment these cubes by resizing them to $512\times 512\times 10$ and applying a {\tt TenCrop} routine to generate $5$ overlapping crops of sizes $256\times 256\times 10$ (positioned at center, top-left, bottom-left, top-right, bottom-right) and their respective vertical flips. 
This yields a total of $2270$ image cubes with final dimensions of $256\times 256\times 10$. 
We use the Focal Tversky Loss \citep[FTL;][]{Abraham2018}) to train the model, which is a generalized version of the Tversky Loss (TL) defined in terms of the Tversky Index (TI)
as,
\begin{equation}
    TI = {TP}/{(TP+\alpha FN + \beta FP)} \rightarrow TL = (1-TI) \rightarrow FTL = (TL)^{\gamma}
\end{equation}
where $TP$, $FN$ and $FP$ are the true positive, false negative and false positive pixels in each slice. The $\gamma$ parameter
guides 
the model to focus on harder ($\gamma<1$) vs. easier ($\gamma>1$) examples. We use $\alpha = 0.7$,  $\beta = 0.3$ and $\gamma=0.7$. For the discriminator optimization, we use the Binary Cross-Entropy ({\it BCE}) loss. 


\section{Results}

\begin{figure}[!h]
    \vspace{-5mm}
    \centering
    \includegraphics[width=\textwidth]{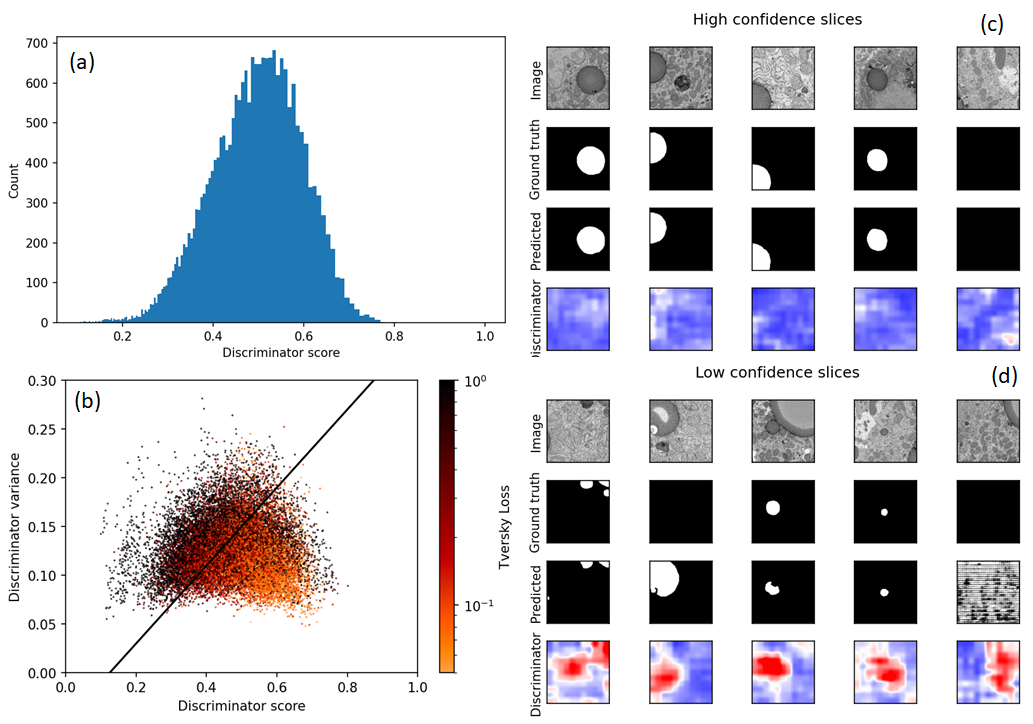}
    \caption{Results from our study. (a) Histogram of mean discriminator scores per slice. (b) Distribution of mean and variance in discriminator score, per slice, 
    with the corresponding segmentation loss. We can use this distribution to make cuts on which slices to show volunteers (region left of the black line). (c, d) Comparison of slices with high discriminator score (more realistic, c) vs low discriminator score (less real, d), with the rows in order showing the input image slice, mask from aggregating volunteers' annotations, the machine predicted segmentation and the patch discriminator output. Blue regions correspond to high discriminator value (more real), while red is low discriminator value (less real). Note how the poorly performing subjects show large feature confusion within the image necessitating further refinement by volunteers. }
    \label{fig:compare}
\end{figure}

The model accuracy, precision and recall are 89.1\%, 49.3\% and 65.4\%, respectively. On Zooniverse, volunteers were asked not to annotate lipid droplets that were not fully visible, but due to our augmentation methods (i.e., cropping of the image cubes) there were cases where partially cropped lipid droplets were shown to the model. As a result, the model learnt to pick up lipid droplets on the edges where volunteers did not annotate any, leading to the reduced precision.

Figure~\ref{fig:compare} presents the test of our proposed framework: panel (a) shows the distribution of discriminator scores for the subjects in our model, panel (b) shows the relation between the discriminator and segmentation loss, and panels (c) and (d) show the comparison of image cubes with high (i.e., more realistic segmentation) and low (i.e., more unrealistic/poorly generalized segmentation) discriminator scores, respectively. 
It is clear that the discriminator has learnt the features corresponding to the lipid droplets (dark circles in the image), and is able to effectively identify where the generator fails to predict these droplets. In fact, the discriminator is able to also show which locations in the image correspond to poor generalization (blue is more real, while red is more fake). The red regions correspond to features that the generator needs to get better at identifying. Images with large red regions can be passed to volunteers to refine the generator's annotation and improve the machine accuracy from re-training.

Figure~\ref{fig:compare}a shows the histogram of discriminator scores for each slice in the dataset (with 1 being very realistic and 0 being unrealistic). We found that while the histogram shows that there are several slices with low discriminator scores (i.e., more unrealistic), there is better separability between slices when we also incorporate the 2D variation in the discriminator score per slice (see Figure~\ref{fig:compare}b). The selection cut described above can the be done by investigating the relation between the mean and the variance in the discriminator score and segmentation loss (Figure~\ref{fig:compare}b). A high discriminator mean corresponds to a more realistic image, while high discriminator variance points to large variation in the generator's performance for that cube. These values are correlated with the segmentation loss $TL$ (i.e., high mean and low variance generally correspond to low segmentation errors), and vice-versa. We can use a selection cut on the discriminator score that would preferentially pick poor performing images and pass it to the volunteers for refinement. For, example, the selection cut proposed in Figure~\ref{fig:compare}b is obtained by choosing a sample of slices with $TL < 0.3$, which produces 9362 slices out of the total of 22700 (a decrease of $64\%$ in the number of images shown to volunteers). 

\section{Conclusions}
We have developed the TCuPGAN, a Convolutional LSTM-based image cube/video segmentation model that simultaneously learns both the 2D spatial and temporal (or depth) signatures of the data. We applied this model on a dataset of microscopy images of the liver tissues from the Etch a Cell - Fat Checker citizen science project on Zooniverse to identify lipid droplets. We find that the adversarial component of the model is effective at identifying slices with poor machine segmentation, and these slices can be provided to the volunteers to correct the machine annotation. Future directions of this work will include deploying a live test project on Zooniverse with this `correct-a-machine' framework to test the increase in project completion rates;
we anticipate that this test will increase the applicability of this model to other use cases on the Zooniverse platform and further reduce the number of images seen by volunteers. 

\bibliographystyle{splncs04}
\bibliography{refs.bib}
\end{document}